\documentclass[journal,10pt,twocolumn,oneside, peereview]{IEEEtran}
\usepackage{subcaption}
\usepackage{enumerate}
\usepackage{float}
\usepackage{amssymb,amsmath,amsxtra,amsfonts,amsthm,mathtools}
\usepackage{graphicx}
\usepackage{bigints}
\usepackage{fancyhdr}
\usepackage{float}
\usepackage{times}
\usepackage{wrapfig}
\usepackage{xcolor,colortbl}
\usepackage{soul}
\usepackage{array,multirow,makecell}

\usepackage{nopageno}
\usepackage{tikz}
\usepackage{glossaries}
\usepackage{hyperref}
\usepackage[switch,columnwise]{lineno} 
\usepackage{mathrsfs}
\usepackage{lineno}
\usepackage[
backend=biber,
style=numeric,
sorting=ynt
]{biblatex}
\usepackage{tikz}
\usetikzlibrary{shapes.geometric, arrows, positioning, fit, calc}

\usepackage{booktabs}
\usepackage{graphicx}
\usepackage{amsthm}

\usepackage{tikz}
\usetikzlibrary{shapes.geometric, arrows}
\usepackage{tikz}
\usetikzlibrary{shapes.geometric, arrows, positioning}

\tikzstyle{startstop} = [rectangle, rounded corners, minimum width=3cm, minimum height=1cm,text centered, draw=black, fill=red!30]
\tikzstyle{io} = [trapezium, trapezium left angle=70, trapezium right angle=110, minimum width=3cm, minimum height=1cm, text centered, draw=black, fill=blue!30]
\tikzstyle{process} = [rectangle, minimum width=3cm, minimum height=1cm, text centered, draw=black, fill=orange!30]
\tikzstyle{arrow} = [thick,->,>=stealth]

\begin{document}
\title{\textbf{Audio-Visual Speech Enhancement: Architectural Design and Deployment Strategies }}

\author{
  Anis Hamadouche,\,
  Haifeng Luo,\,\,
  Mathini Sellathurai,\,\,
  Amir Hussain\,\,
  and\,
  Tharm Ratnarajah%
  \IEEEcompsocitemizethanks{
    \IEEEcompsocthanksitem
    Work supported by the UK Engineering and Physical Sciences Research Council (EPSRC) Grant No.  EP/T021063/1. 
    \IEEEcompsocthanksitem
        Anis Hamadouche and Mathini Sellathurai
    are with the School of Engineering \& Physical Sciences, Heriot-Watt University, Edinburgh EH14 4AS,
    UK (e-mail: \{anis.hamadouche,m.sellathurai\}@hw.ac.uk).  Tharmalingam Ratnarajah is with the  College of Engineering Department of Electrical and Computer Engineering, San Diego State University, USA (e-mail:  tratnarajah@sdsu.edu).
    Amir Hussain is with SDAIA-KFUPM Joint Research Centre for Artificial Intelligence, King Fahd University of Petroleum and Minerals, Dhahran, Saudi Arabia (e-mail: amir.hussain@kfupm.edu.sa).
  }
}

\maketitle
\thispagestyle{plain}
\pagestyle{plain}

\begin{abstract}
\textbf{Real-time audio-visual speech enhancement (AVSE) is a key enabler for immersive and interactive multimedia services, yet its performance is tightly constrained by network latency, uplink capacity, and computational delay. This paper presents the design, deployment, and evaluation of a complete cloud–edge–assisted AVSE system operating over a public 5G edge network. The system integrates CNN-based acoustic enhancement and OpenCV-based facial feature extraction with an LSTM fusion network to preserve temporal coherence, and is deployed on a Vodafone-compatible AWS Wavelength edge cloud. Through extensive stress testing, we analyze end-to-end performance under varying network load and adaptive multimedia profiles. Results show that compute placement at the network edge is critical for meeting real-time coherence constraints, and that uplink capacity is often the dominant bottleneck for interactive AVSE services. Only 5G and wired Ethernet consistently satisfied the required communication delay bound for uncompressed audio-video chunks, while aggressive compression reduced payload sizes by up to 80× with negligible perceptual degradation, enabling robust operation under constrained conditions. We further demonstrate a fundamental trade-off between processing latency and enhancement quality, where reduced model complexity lowers delay but degrades reconstruction performance in low-SNR scenarios. Our findings indicate that public 5G edge environments can sustain real-time, interactive AVSE workloads when network and compute resources are carefully orchestrated, although performance margins remain tighter than in dedicated infrastructures. The architectural insights derived from this study provide practical guidelines for the design of delay-sensitive multimedia and perceptual enhancement services on emerging 5G edge-cloud platforms.}
\end{abstract}

\IEEEpeerreviewmaketitle
\newcommand\scalemath[2]{\scalebox{#1}{\mbox{\ensuremath{\displaystyle #2}}}}

\section{Introduction}

Speech signals captured in real-world environments are often severely degraded by background noise, reverberation, and competing speakers, leading to reduced intelligibility and perceived quality. Traditional audio-only speech enhancement techniques struggle to recover clean speech in low signal-to-noise ratio (SNR) settings or with overlapping talkers, because audio cues alone provide limited information under such conditions. In contrast, humans naturally exploit visual cues—such as lip movements and facial articulations—to disambiguate speech in noise, demonstrating that visual information is robust to acoustic masking and can significantly improve speech understanding in adverse environments. Audio-visual speech enhancement (AVSE) systems leverage this multimodal insight by fusing acoustic and visual features through deep neural networks, often outperforming audio-only approaches in low-SNR or multi-speaker scenarios.

Despite these advances in multimodal modeling, deploying AVSE in real-time applications introduces a distinct set of system-level challenges. Wearable and mobile devices—such as hearing aids, smart glasses, or assistive communication tools—must operate under strict latency, power, and connectivity constraints while processing continuous streams of synchronized audio and video. Offloading compute-intensive deep-learning inference to remote servers emerges as a viable solution for resource-limited devices, but this exposes the system to network latency, jitter, and variable throughput that can degrade performance or violate temporal coherence requirements.

The advent of 5G and edge cloud computing offers a promising platform for addressing these challenges. Multi-access edge computing (MEC) brings cloud capabilities closer to end users by deploying compute resources at or near the radio access network (RAN), thereby reducing data transport delay and backbone congestion. Standards bodies such as ETSI define MEC as an environment where applications can run at the edge of 4G/5G networks with ultra-low latency and high bandwidth~\cite{etsi_mec}. Edge cloud platforms like AWS Wavelength embed cloud infrastructure within telecommunications networks, enabling low-latency access to compute and storage services directly from 5G devices~\cite{aws_wavelength}. While prior work has demonstrated the theoretical potential of MEC for latency-sensitive services, real-world empirical studies of audio-visual multimedia workloads over public 5G edge deployments remain limited, especially at the intersection of multimodal processing, network variability, and end-to-end performance.

Audio-visual multimedia workloads are inherently demanding: they combine temporally synchronized audio and video streams, require sustained throughput for high-resolution visual data, and must maintain tight end-to-end latency to preserve perceptual coherence. Even modest delay variation or packet loss can disrupt synchronization, degrade quality, and undermine user experience in interactive settings. These challenges are further exacerbated in mobile environments, where wireless channel conditions and user mobility introduce variability that centralized cloud architectures cannot easily accommodate.

In this work, we address these gaps by presenting a full-stack, deployment-oriented study of audio-visual multimedia delivery and processing over a public 5G edge cloud, using AVSE as a representative real-time workload. We design and implement an AVSE pipeline that captures synchronized audio and lip-region video on a wearable front end, transmits data via a 5G uplink to an edge cloud server, performs multimodal deep learning-based enhancement, and streams enhanced speech back to the device. The system is deployed on a commercial 5G edge platform, and its performance is evaluated in terms of end-to-end latency, throughput, packet loss, compute utilization, and enhancement quality under realistic network conditions and adaptive stress profiles.

Our contributions are as follows:

\begin{itemize}
\item We present a real-world deployment and end-to-end evaluation of audio-visual multimedia processing over a public 5G edge cloud, extending AVSE research from model-centric prototypes to practical systems.
\item We introduce stress-adaptive multimedia profiles that systematically explore throughput, latency, and quality trade-offs across heterogeneous network conditions.
\item We quantitatively measure network and system performance—including uplink throughput, round-trip latency, resource utilization, and packet loss—to identify constraints on real-time operation in public 5G environments.
\item We derive practical insights and design guidelines for latency-sensitive multimedia services on 5G edge clouds, highlighting the importance of edge proximity, adaptive profiles, and multi-modal synchronization for real-time QoE.
\end{itemize}

The remainder of this paper is organized as follows. Section II reviews related work on audio-visual speech enhancement, edge/cloud deployment strategies, and 5G-enabled multimedia systems. Section III presents the proposed audio-visual speech enhancement architecture and its deployment over a public 5G edge cloud, including system design and latency modeling. Section IV reports experimental results from real-world deployments over Ethernet, Wi-Fi, 4G, and public 5G networks, and analyzes throughput, latency, reliability, and compute utilization under stress-adaptive multimedia profiles. Section V discusses algorithmic processing latency and model optimization for real-time operation. Finally, Section VI concludes the paper and outlines directions for future research.

\section{Related Work}

Research on audio–visual speech enhancement (AVSE) has advanced significantly with the adoption of deep learning and multimodal fusion architectures. Early seminal work demonstrated that combining audio with visual cues such as lip motion significantly improves speech separation and enhancement under noisy and multi-speaker conditions compared to audio-only methods. For example, Ephrat \emph{et al.} jointly model audio and visual modalities for speaker separation, establishing a foundation for multimodal speech processing~\cite{ephrat2018looking}. Subsequent models such as multimodal deep convolutional networks (e.g., AVDCNN) validate the effectiveness of integrated audio and visual feature extraction and joint decoding for enhanced speech quality~\cite{hou2018audio}.

Beyond model design, more recent AVSE studies have explored enhanced fusion strategies (e.g., dense connectivity and gated fusion) and real-time capable architectures, while also considering robust dataset construction and domain-specific tasks such as spatial audio processing or feature alignment~\cite{zhu2023real}.

While the majority of AVSE research focuses on improving model performance in controlled conditions, fewer studies tackle the challenges of real-time deployment where network transport, latency constraints, and computational placement are critical. In~\cite{gupta20235g} the authors showcased a proof-of-concept implementation of the world’s first 5G and Internet of Things (IoT) enabled multi-modal hearing aid (MM HA) prototype. This integrates an innovative 5G cloudradio access network (C-RAN) and IoT based transceiver model for real-time audio-visual speech enhancement (AVSE). Our work evaluates AVSE performance under realistic network conditions (Ethernet, Wi-Fi, 4G, and 5G), highlighting trade-offs between communication delay, model complexity, and enhancement effectiveness.

Another related direction investigates 5G-enabled IoT and wearable systems, where audiovisual data are streamed to cloud servers for enhancement and secure processing, addressing both latency and privacy concerns in emerging assistive applications~\cite{adeel2018real}. Beyond AVSE, broader work on mobile edge computing (MEC) and 5G performance analysis provides insights into system design for interactive multimedia services, where computation offloading and resource allocation strategies are critical to support low-latency processing required by live audiovisual applications~\cite{mao2017survey}. Additionally, work on wireless offloading and wearable edge computing suggests that high-throughput, low-latency links can effectively support the real-time transport of high-resolution visual streams for vision-driven workloads such as object detection, further emphasizing the potential of advanced wireless architectures for AVSE systems~\cite{yuan2022network}.





\section{Problem Formulation: Audio–Visual Speech Enhancement}

Audio–visual speech enhancement (AVSE) aims to recover a clean target speech signal by jointly exploiting
acoustic and visual observations of the speaker. Let the discrete-time noisy audio signal captured at the
microphone be
\begin{equation}
y[n] = x[n] + d[n],
\end{equation}
where $x[n]$ denotes the clean target speech and $d[n]$ represents additive noise and interference.
In adverse acoustic environments, recovering $x[n]$ from $y[n]$ alone is ill-posed, particularly at low
signal-to-noise ratio (SNR) or in the presence of competing speakers. Visual cues, however, remain largely
invariant to acoustic corruption and therefore provide complementary information for speech recovery.

Let $\mathbf{a}_t \in \mathbb{R}^{F_a}$ denote the audio feature vector extracted from $y[n]$ over time
frame $t$ (e.g., log-mel or STFT magnitude features), and let $\mathbf{v}_t \in \mathbb{R}^{F_v}$ denote the
corresponding visual feature vector extracted from the speaker’s facial region (e.g., lip motion
embeddings). The AVSE problem is formulated as learning a nonlinear mapping
\begin{equation}
\hat{\mathbf{x}}_t = f_\theta(\mathbf{a}_{1:t}, \mathbf{v}_{1:t}),
\end{equation}
where $f_\theta(\cdot)$ denotes a deep neural network with parameters $\theta$, and
$\hat{\mathbf{x}}_t$ is the estimated clean speech representation at time $t$.
The model is trained by minimizing the reconstruction loss
\begin{equation}
\min_{\theta} \; \mathcal{L}
= \sum_{t=1}^{T} \left\| \hat{\mathbf{x}}_t - \mathbf{x}_t \right\|_2^2,
\end{equation}
where $\mathbf{x}_t$ denotes the clean target speech features and $T$ is the sequence length.
\section{Proposed Solution}
\label{Proposed Solution}

\subsection{Model Architecture and Feature Fusion}

\begin{figure}[h!]
    \centering
    \includegraphics[width=\columnwidth]{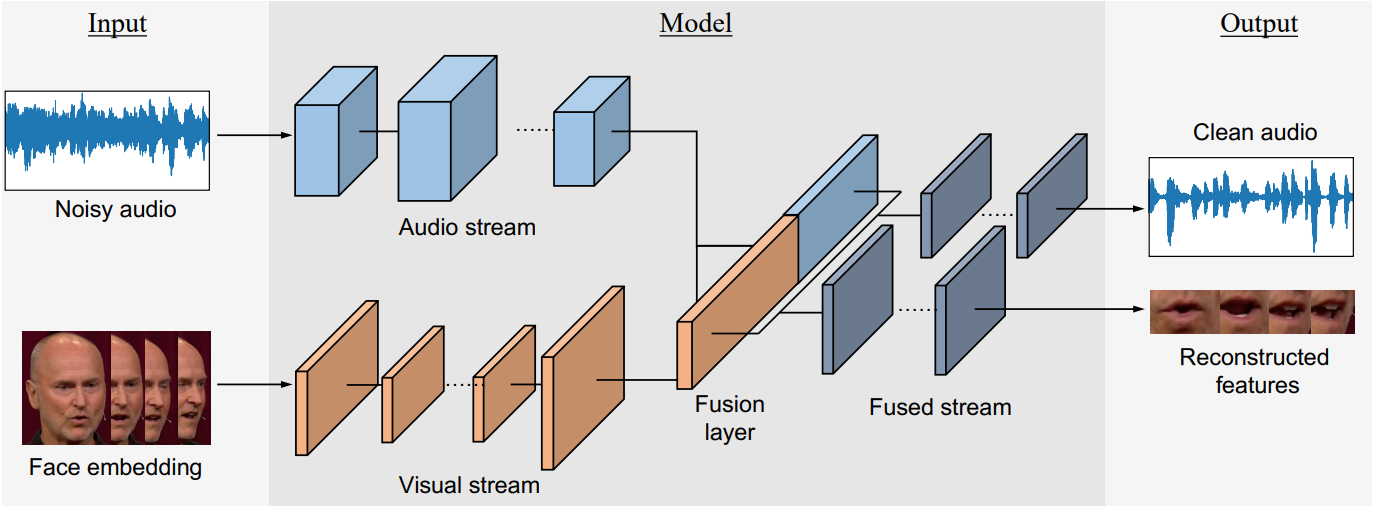}
    \caption{Audio-Visual Speech Enhancement AI model architecture.}
    \label{fig:avse_arch}
\end{figure}

The proposed AVSE architecture consists of parallel audio and visual processing streams followed by
feature fusion and temporal modeling, as illustrated in Fig.~\ref{fig:avse_arch}.

The audio stream employs a convolutional neural network (CNN) encoder
\begin{equation}
\mathbf{z}_t^{(a)} = \mathcal{E}_a(\mathbf{a}_t),
\end{equation}
while the visual stream extracts facial features using a vision front-end based on OpenCV and CNN
feature extractors,
\begin{equation}
\mathbf{z}_t^{(v)} = \mathcal{E}_v(\mathbf{v}_t).
\end{equation}
The modality-specific embeddings are fused via concatenation,
\begin{equation}
\mathbf{z}_t = \left[ \mathbf{z}_t^{(a)} \; \| \; \mathbf{z}_t^{(v)} \right],
\end{equation}
where $[\cdot \| \cdot]$ denotes vector concatenation. Temporal dependencies are modeled using a long
short-term memory (LSTM) network,
\begin{equation}
\mathbf{h}_t = \mathrm{LSTM}(\mathbf{z}_t, \mathbf{h}_{t-1}),
\end{equation}
which ensures coherent speech reconstruction across time. A stack of fully connected (FC) layers
decodes the enhanced speech features,
\begin{equation}
\hat{\mathbf{x}}_t = \mathcal{D}(\mathbf{h}_t).
\end{equation}




\subsection{Implementation}

\begin{figure*}
    \centering
    \includegraphics[width=\textwidth]{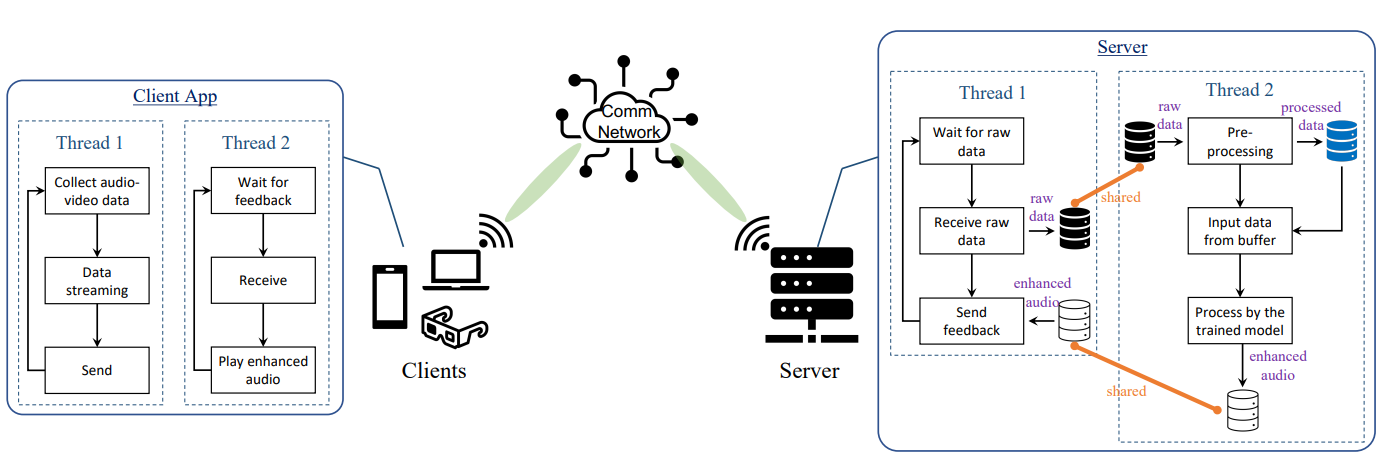}
    \caption{A block diagram of enabling real-world AVSE service on terminal devices.}
    \label{figure2}
\end{figure*}

\autoref{figure2} presents a pragmatic framework for the deployment of the developed AVSE algorithm within the context of real-world applications. In order to enable the service in low quality terminals (e.g., IoT devices), the remote server takes over all the processing, including the pre-processing of the raw data. By doing so, any device with a camera, microphone, and communication capabilities can use the service without any requirements on its computing power. This allows us to apply the service to some cheap IoT devices to achieve specific functions. For high-performance devices (e.g., smartphones), offloading all processing to the server can also reduce energy consumption and improve battery life.

As the figure depicts, the client terminals only collect the raw audio and video, then upload them to the server and wait for the feedback from the server. It should be noted that collecting and playing audio requires parallel processing, so we use two independent threads. The server will listen to clients and wait for the raw audio-video data from clients. It starts processing using the stored trained model once it listens to data transmission from the client. Considering that processing latency may exceed the duration of the audio itself, resulting in incoherent feedback audio, we use buffers to store input and output data and perform parallel signal processing in another independent thread. In this way, even if the processing delay is high due to algorithm complexity, the client can still hear coherent enhanced audio after a specific delay, thereby having a satisfactory service experience.

\begin{figure}
    \centering
    \includegraphics[width=\columnwidth]{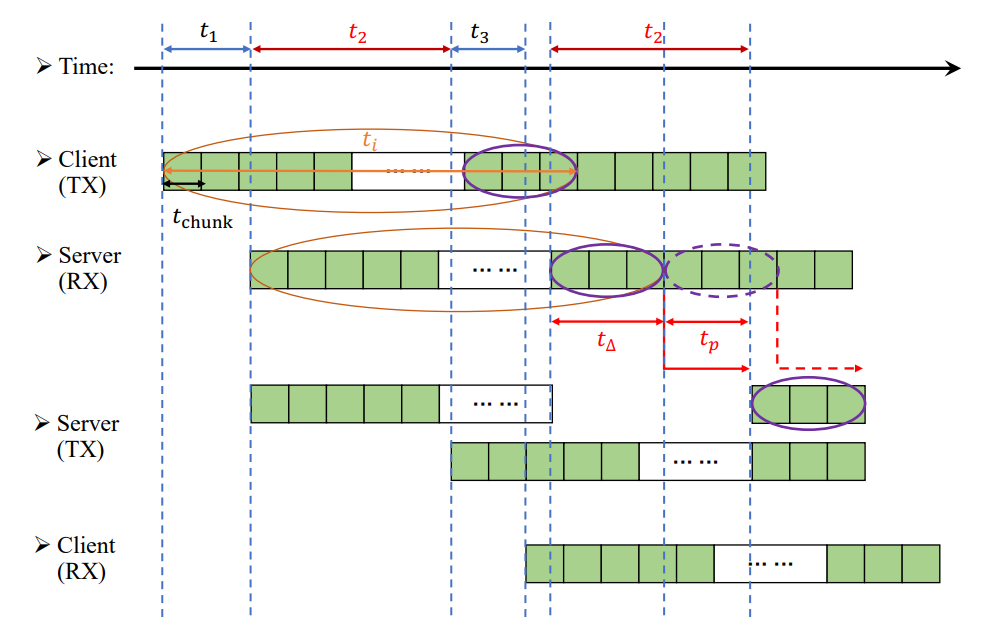}
    \caption{A schematic describes the experienced latency of this service.}
    \label{figure3}
\end{figure}

\autoref{figure3} describes the latency of the processing of the AVSE service. As shown in the system model, the client continuously collects voice and video information and sends it to the server. Assume that the length of the audio-video chunk sent by the client each time is $t_{\text{chunk}}$, that is, the client will send out audio and video after collecting a chunk of $t_{\text{chunk}}$. After receiving the data, the server will first perform preprocessing, such as feature extraction of video frames, and then store the processed data in the buffer. Since our algorithm tracks lips to help audio enhancement, the input of the algorithm must have a specific length $t_i$ to ensure sufficient visual features. When there are more than $t_i$ data in the buffer, the server will start executing the AVSE algorithm. Then, the server processes the incoming audio and video data sequentially, and we must set a sufficient buffer interval $t_{\Delta}$ to ensure that it can complete each process, which is given as

\begin{equation}
    t_{\Delta} > t_a
\end{equation}

where $t_a$ is the algorithm processing latency. The enhanced audio is stored in the output buffer, and the server begins feeding the enhanced audio back to the client when there is enough data in the output buffer. It should be noted that since  $t_i$ is usually much larger than $t_{\Delta}$, if we want the initial audio to be enhanced by this algorithm, there will be an inevitable $t_i$-length waiting time for data, causing a high delay. In order to mitigate the significant delays we do not process the audio at the beginning of  $t_i - t_{\Delta}$, thereby reducing the overall delay to

\begin{equation}
    t_{\text{delay}} = t_1 + t_2 + t_3 = t_{\text{comm}} + t_{\text{proc}}
\end{equation}

where $t_1$ denotes the forward network latency, i.e., the delay incurred in transmitting data from the client to the server; $t_2$ denotes the server’s processing latency, i.e., the delay incurred by the server in processing the data; $t_3$ denotes the reverse network latency, i.e., the delay incurred in transmitting data from the server to the client. Thus, $t_1+ t_3$ comprise the communication latency, and the processing latency $t_{\text{proc}}$ (i.e., $t_2$) can be decomposed into the buffer interval $t_{\Delta}$ and algorithm processing time $t_a$. As a result, the client can get coherent feedback audio $t_{\text{proc}}$ seconds after enabling the service, and audio enhancement is enabled after  $t_i - t_{\Delta}$ seconds.

\begin{figure}[t]
    \centering
    \includegraphics[width=\columnwidth]{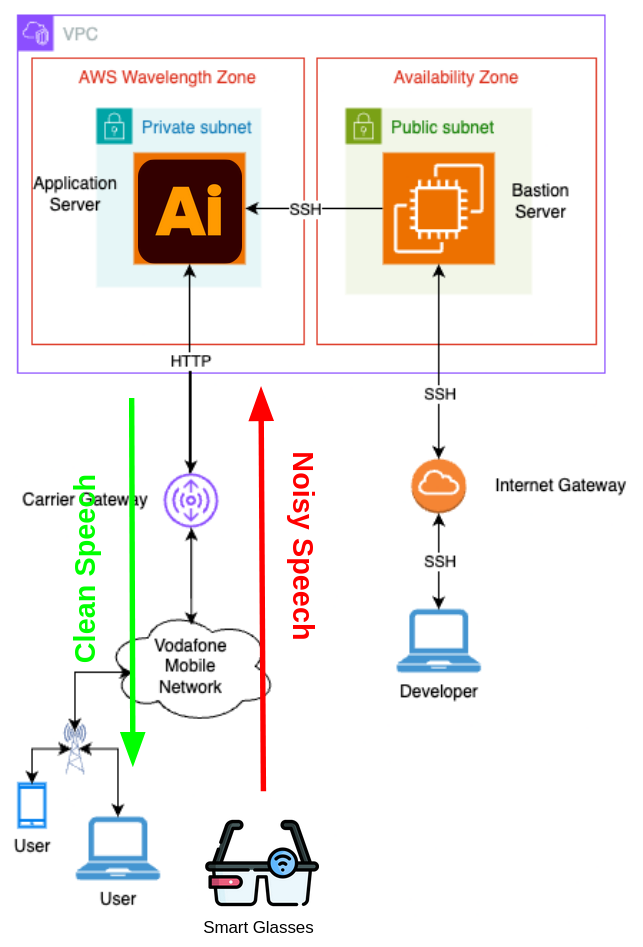}
    \caption{End-to-end system architecture for 5G edge cloud multimedia delivery.}
    \label{fig:architecture}
\end{figure}

\section{Experimental Results}
\label{Experimental Results}

In this experiment, the auditory information undergoes a transformation via convolutional neural networks (CNNs), which can distill the audio patterns. The visual information is processed through the OpenCV library, which is an open-source computer vision and machine learning software library, to extract facial features from the face embeddings. It is implemented using the pre-trained Haar feature-based cascade classifiers, which are underpinned also by CNNs. Subsequently, the algorithm employs a fusion strategy where the extracted auditory and facial features coalesce, serving as inputs to a sophisticated Long- Short Term Memory (LSTM) network. This step is crucial for integrating temporal dynamics and ensuring a coherent feature representation. The convergence of these modalities culminates in a series of FC layers, which tunes the composite features to reconstruct the target outputs. This framework provides a foundational blueprint, while the explicit architectural configuration (i.e., layers, neurons, etc.) can vary depending on the requirements and data pre-processing.

It should be noted that although we do buffering to ensure coherent audio feedback from the server, the capacity of the communication network is also extremely important to ensure reliable quality of service. The communication network needs to have sufficient capacity so that the round-trip delay of data transmission between the client and server is shorter than the time of the data itself to ensure a consistent service experience, denoted as 

\begin{equation}
    t_{\text{comm}} \leq t_{\text{chunk}}
\end{equation} 

Otherwise, after playing a chunk, if the data of the next chunk has not been fed back, there will be a blank period. 

\begin{figure*}[t]
    \centering
    
    \begin{minipage}{0.32\textwidth}
        \centering
        \includegraphics[width=\textwidth]{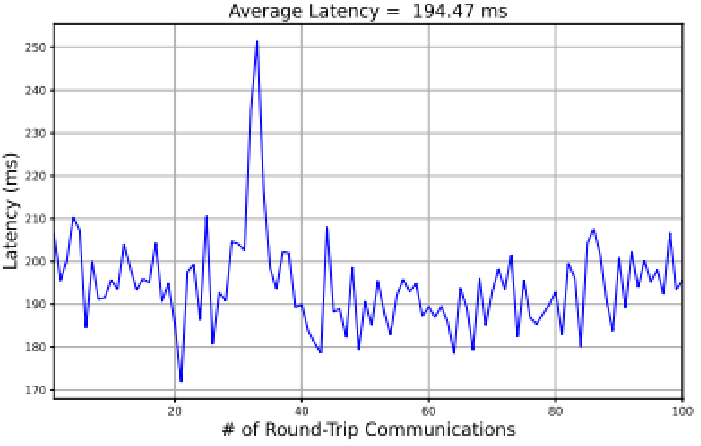}
        \caption*{(a) Wi-Fi (802.11n)}
        \label{fig:latency_wifi}
    \end{minipage}
    \hfill
    \begin{minipage}{0.32\textwidth}
        \centering
        \includegraphics[width=\textwidth]{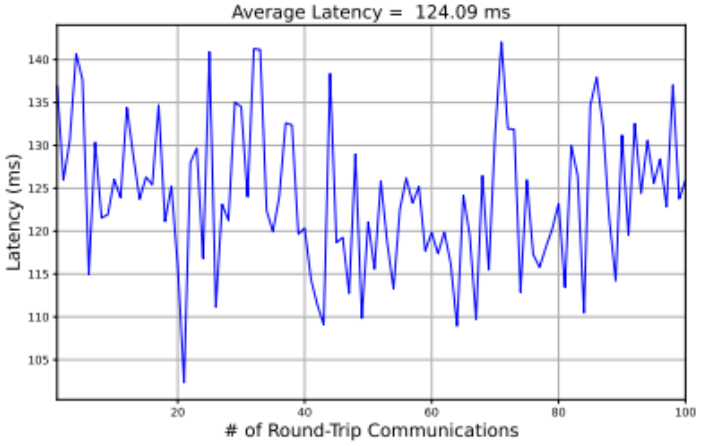}
        \caption*{(b) 4G}
        \label{fig:latency_4g}
    \end{minipage}
    \hfill
    \begin{minipage}{0.32\textwidth}
        \centering
        \includegraphics[width=\textwidth]{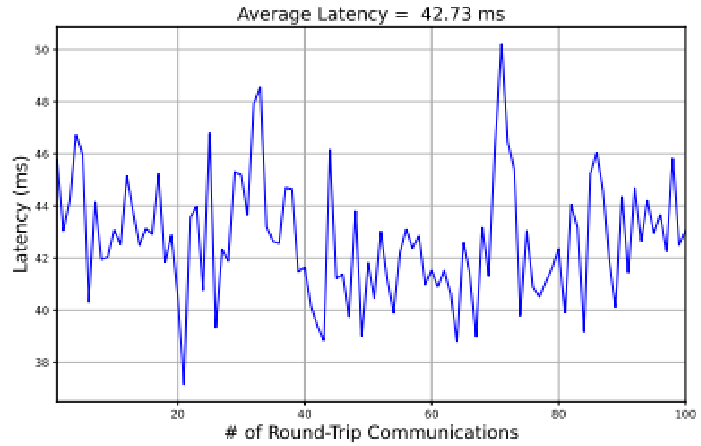}
        \caption*{(c) 5G}
        \label{fig:latency_5g}
    \end{minipage}

    \vspace{0.3cm}

    \begin{minipage}{0.32\textwidth}
        \centering
        \includegraphics[width=\textwidth]{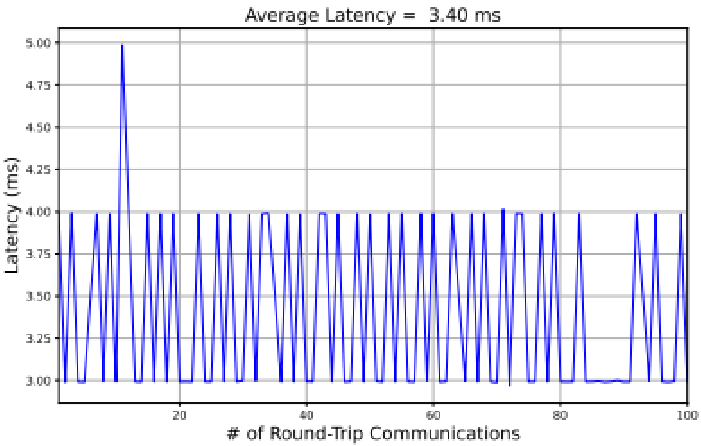}
        \caption*{(d) Ethernet cable}
        \label{fig:latency_ethernet}
    \end{minipage}
    \hfill
    \begin{minipage}{0.32\textwidth}
        \centering
        \includegraphics[width=\textwidth]{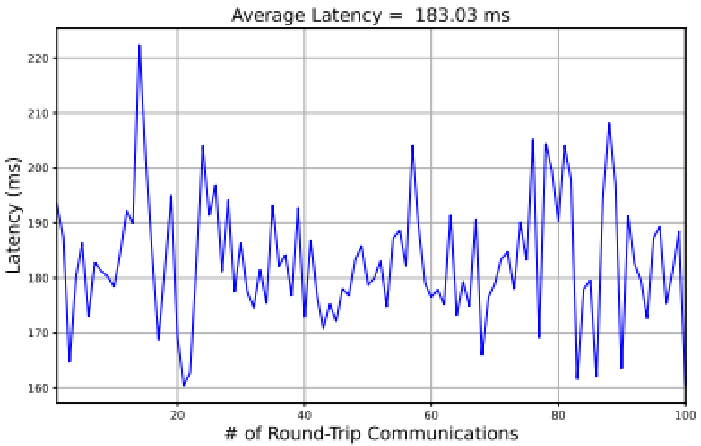}
        \caption*{(e) Amazon Web Services}
        \label{fig:latency_aws}
    \end{minipage}
    \hfill
    \begin{minipage}{0.32\textwidth}
        \centering
        \includegraphics[width=\textwidth]{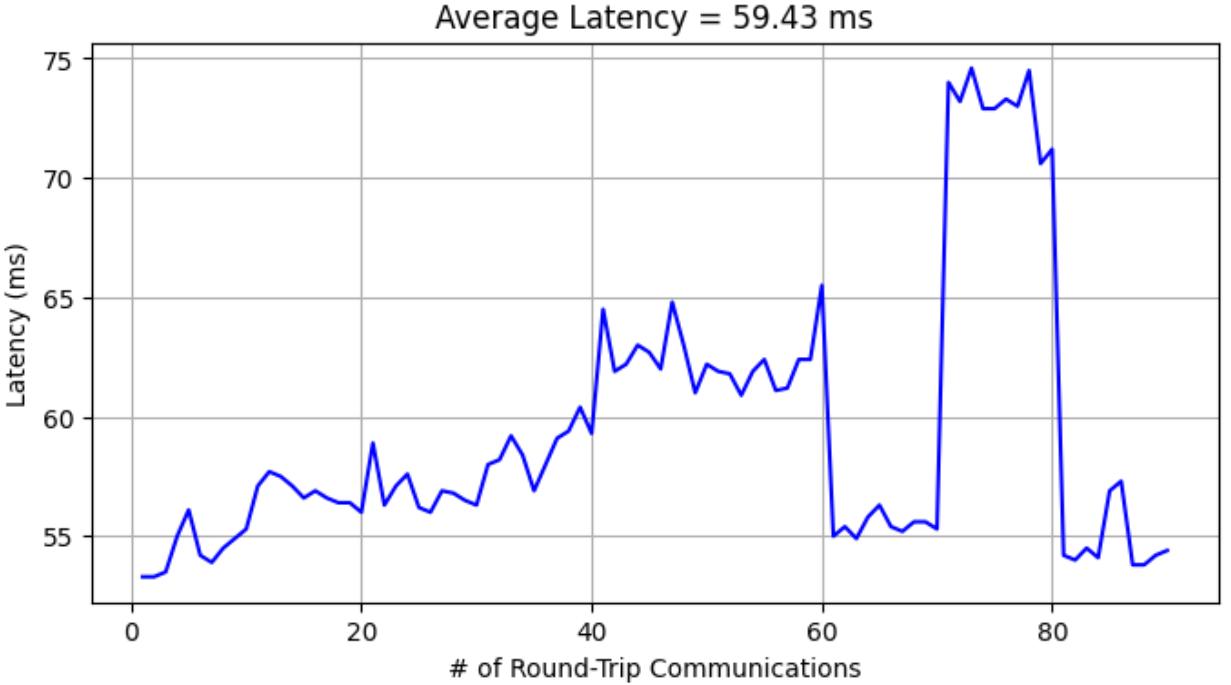}
        \caption*{(f) AWS Wavelength over Vodafone 5G Cellular Network}
        \label{fig:latency_wavelength_vodafone}
    \end{minipage}

    \caption{Network latency measured in real-world environments (round-trip latency for transferring $\tilde 0.3$ MB of data).}
    \label{figure4}
\end{figure*}

We measure the latency of a variety of networks in real-world environments. We record audio at a sampling rate of $16$ kHz and captured video frames at a resolution of $380\times640$ pixels at a frame rate of $25$ fps. We set $t_{\text{chunk}} = 40$ ms, congruent with the temporal extent of a single video frame. This configuration yields a raw data payload of approximately $0.3$ MB per audio-video chunk. \autoref{figure4} shows the latency profiles observed across a suite of $100$ round-trip data transmissions. In our experimental setup, a desktop-based server and a laptop-based client were utilized to simulate a realistic user scenario. The communication between server and client was facilitated through Wi-Fi (802.11n, i.e., Wi-Fi 4), 4G, 5G, and a direct Ethernet cable link, respectively. We also tried the Amazon Web Services over internet as well as over Vodafone 5G public network whose architecture is depicted in~\autoref{fig:architecture}. The results suggest that only private 5G network and direct Ethernet cable link could support a coherent service experience, i.e., satisfying $t_{\text{comm}} \leq t_{\text{chunk}}$ where $t_{\text{chunk}} = 40$ ms in our experiments. Note that 4G and 5G network capacities may differ in different regions and with different carriers while choosing different AWS locations may also result in different latencies. We used Wi-Fi from the King’s Buildings, The University of Edinburgh (Edinburgh, UK), and for 4G and 5G network testing used a private network of Cambridge Wireless (Cambridge, UK) and Vodafone 5G public network. Using new Wi-Fi protocols, such as 802.11ac (Wi-Fi 5) or 802.11ax (Wi-Fi 6), can dramatically reduce the latency. 

\begin{figure}
    \centering
    \includegraphics[width=\columnwidth]{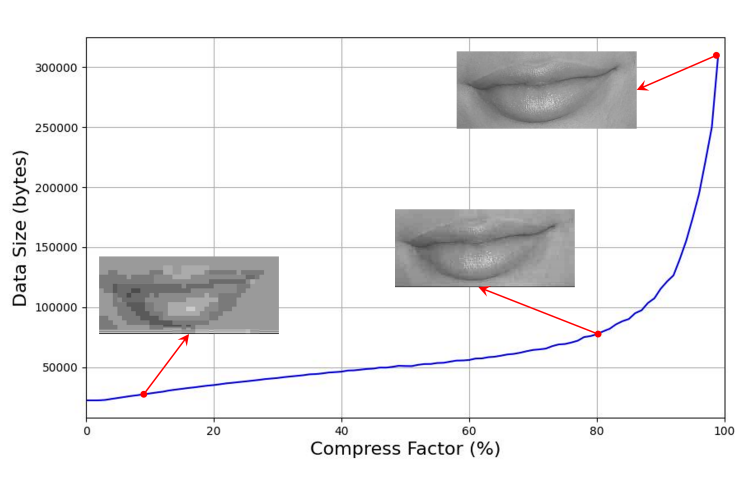}
    \caption{Audio-video data size versus compression factors.}
    \label{figure5}
\end{figure}

In areas with poor signal coverage, we can compress the raw audio-video data to reduce the transmission load, obtaining reliable and consistent quality of service. \autoref{figure5} shows the data size versus compression factors. It can be seen that proper compression of data can greatly reduce the data size and the required data rate, especially in the range of 80-100 compression factors. At a compression factor of 80, the data can be reduced to $\approx 70$ KB with an insignificant effect on the quality of the video frames. We also used a compression factor of 80 in our real-world demonstration in Cambridge to ensure smooth service, and experiments proved that the impact of other speech enhancement properties could be ignored and we could still achieve considerable noise reduction.


The next section describes extensive Vodafone 5G public network stress testing with different data compression profiles.

\subsection{Vodafone 5G network}
\label{sec:exp_setup}

This section describes the experimental environment used to evaluate real-time audio-visual multimedia delivery over a public 5G network integrated with edge cloud infrastructure. We first outline the network characteristics and edge cloud configuration, followed by details of the user equipment setup, stress profile design, and performance metrics used for analysis.

The experiments were conducted over a commercial Vodafone 5G network operating in the United Kingdom’s mid-band spectrum (3.4--3.6\,GHz), which provides a balance of capacity and coverage suitable for real-world edge deployments. Targeted network characteristics include an end-to-end latency of less than 15\,ms for edge-hosted applications and sustained upstream bandwidth in excess of 50\,Mbps. Typical downstream 5G throughput is expected to range between 100--200\,Mbps, significantly higher than legacy 4G under ideal conditions, thus accommodating bandwidth-intensive audio-visual workloads.

The edge cloud environment was hosted using AWS Wavelength Zones, which embed compute and storage resources directly within the operator’s network to minimize transport distance. In this deployment, application servers were placed within a private subnet accessible only through the operator’s carrier gateway. A bastion host provided controlled secure shell access for development and system management, ensuring that compute resources remained isolated from direct public Internet exposure. General-purpose instance types supported by Wavelength include t3.medium, t3.xlarge, and r5.2xlarge, along with GPU-accelerated options such as g4dn.2xlarge; this work uses the t3.medium instance type to represent a typical baseline for edge compute.

User equipment consisted of a Google Pixel~8 smartphone used as the primary 5G access device. To facilitate logging and control from a standard development environment, the phone’s mobile data connection was shared with a laptop via Wi-Fi tethering, effectively creating a local hotspot. All multimedia payloads were generated by client applications on the tethered laptop and transmitted over the 5G access network to the edge application servers. This setup ensured realistic traffic behavior on the mobile access link while providing flexibility for measurement and control.

To systematically explore system performance under varying network loads and processing demands, we defined a set of AVSE profiles. These profiles vary in video frame rate, spatial resolution, JPEG compression quality, and audio frame configuration, reflecting a range from very low to extreme bandwidth usage. Table~\ref{tab:profiles} summarizes the profile parameters used in this study.

\begin{table*}[t]
\centering
\caption{AVSE Multimedia Profile Configurations}
\label{tab:profiles}
\begin{tabular}{c l c c c c l}
\toprule
\textbf{Profile} & \textbf{Category} & \textbf{Video FPS} & \textbf{Resolution} & \textbf{JPEG Quality} & \textbf{Audio Frame (ms)} & \textbf{Bandwidth Class} \\
\midrule
1A & Low-bandwidth & 3 & 160$\times$120 & 60 & 20 & Very low \\
1B & Low-bandwidth & 5 & 240$\times$180 & 70 & 20 & Low \\
2A & Medium-bandwidth & 7 & 320$\times$240 & 75 & 20 & Medium \\
2B & Medium-bandwidth & 10 & 320$\times$240 & 90 & 20 & Medium–high \\
3A & High-bandwidth & 10 & 640$\times$480 & 80 & 20 & High \\
3B & High-bandwidth & 15 & 640$\times$480 & 85 & 20 & Very high \\
4A & Stress-test & 20 & 320$\times$240 & 70 & 20 & High (FPS-heavy) \\
4B & Stress-test & 30 & 1280$\times$720 & 95 & 20 & Extreme \\
5A & Audio-only & -- & -- & -- & 20 & Very low \\
5B & Audio-only & -- & -- & -- & 10 & Low \\
\bottomrule
\end{tabular}
\end{table*}

Performance was assessed using several key metrics. Round-trip time (RTT) was measured between the user equipment and the edge application server, excluding application processing overhead, to characterize pure network latency. Throughput measurements captured sustained uplink and downlink bandwidth during multimedia transmission. Packet loss was tracked at the network level to quantify reliability, and jitter was analyzed to understand timing variation affecting real-time media. Finally, resource utilization on the edge compute node, particularly host CPU usage, was monitored to evaluate processing overhead under different profile loads. Together, these metrics provide a comprehensive view of network, transport, and system performance supporting real-time multimedia over 5G edge cloud deployments.

The analysis below focuses on throughput scalability, latency behavior, system resource utilization, and reliability across AVSE profiles ranging from audio-only baselines to extreme audio-visual configurations.


\begin{figure*}[t]
    \centering

    \begin{subfigure}[b]{0.48\textwidth}
        \centering
        \includegraphics[width=\columnwidth]{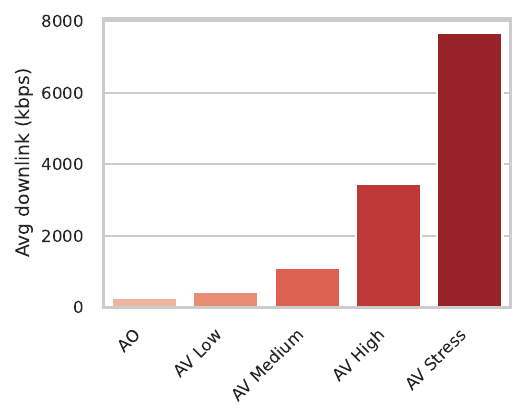}
        \caption{Average downlink throughput.}
        \label{fig:throughput}
    \end{subfigure}
    \hfill
    \begin{subfigure}[b]{0.48\textwidth}
        \centering
        \includegraphics[width=\columnwidth]{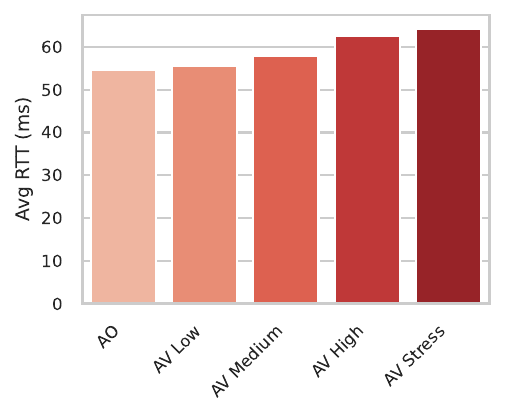}
        \caption{Average RTT (network latency only).}
        \label{fig:rtt}
    \end{subfigure}

    \vspace{0.5em}

    \begin{subfigure}[b]{0.48\textwidth}
        \centering
        \includegraphics[width=\columnwidth]{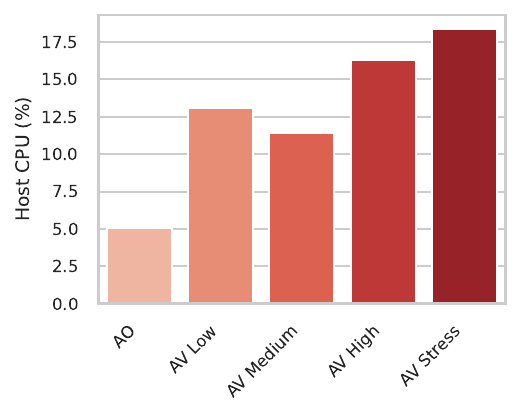}
        \caption{Average host CPU utilization.}
        \label{fig:cpu}
    \end{subfigure}
    \hfill
    \begin{subfigure}[b]{0.48\textwidth}
        \centering
        \includegraphics[width=\columnwidth]{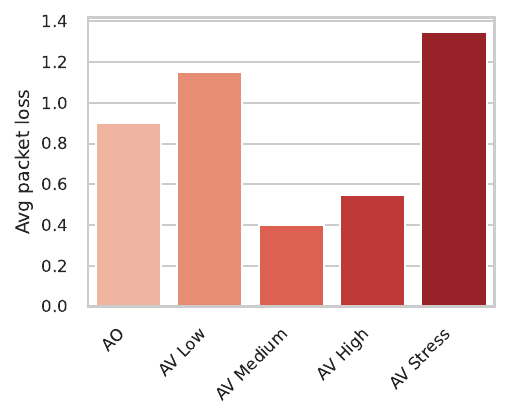}
        \caption{Average packet loss.}
        \label{fig:loss}
    \end{subfigure}

    \caption{AWS edge-cloud (Wavelength) over Vodafone 5G network performance metrics across different AVSE profiles under increasing load.}
    \label{fig:stress_metrics}
\end{figure*}


Fig.~\ref{fig:throughput} presents the average downlink throughput for the five stress classes. Throughput increases monotonically from the audio-only (AO) baseline to the AV Stress profile, demonstrating that the proposed profile design provides controlled and predictable scaling of network load. Low and medium stress profiles (AV Low and AV Medium) increase throughput gradually, enabling graceful adaptation under constrained network conditions. In contrast, the AV High and AV Stress profiles introduce a sharp increase in bandwidth demand, with the latter exceeding 13~Mbps on average, representing the upper operational limit of the pipeline. These results validate the ability of the system to span a wide range of deployment scenarios using a single unified profile design.



Fig.~\ref{fig:rtt} shows the corresponding network latency measured as average RTT, excluding AVSE processing overhead. RTT remains relatively stable for AO, AV Low, and AV Medium profiles (approximately 54--59~ms), indicating that moderate increases in video bitrate do not significantly impact network delay when edge cloud deployment is used. However, latency increases for AV High and peaks for AV Stress at approximately 73~ms, suggesting the onset of queuing and scheduling effects under sustained high throughput. While still acceptable for real-time audio-visual processing, these results highlight the reduced latency margin at extreme configurations and the importance of adaptive profile selection.



Fig.~\ref{fig:cpu} illustrates the host CPU utilization across stress levels. Audio-only processing requires minimal resources (approximately 5\% CPU usage), while AV Low and AV Medium profiles operate between 10--13\%, representing a practical region for sustained deployment. The AV High profile increases CPU usage to approximately 18\%, and the AV Stress profile exceeds 21\%, reflecting the combined cost of high-rate video encoding and network transmission. These results indicate that computational load scales predictably with stress level and that mid-range profiles provide the best efficiency–performance trade-off.



Fig.~\ref{fig:loss} reports the average packet loss observed across stress profiles. Packet loss remains low for AO through AV High profiles, demonstrating the robustness of the pipeline under typical operating conditions. The AV Stress profile exhibits higher loss and variance, confirming that it is primarily suitable for stress testing rather than continuous real-time operation. Notably, AV Medium and AV High profiles maintain low packet loss while delivering significantly higher throughput, reinforcing their suitability for practical deployments.


Overall, the results demonstrate that while extreme profiles expose system and network limits, mid-range configurations achieve a favorable balance between throughput, latency, resource consumption, and reliability. These findings confirm that edge cloud proximity, combined with adaptive profile selection, is a key enabler for real-time audio-visual speech enhancement in public 5G environments.

\begin{table*}[t]
\centering
\caption{Latency report of different AVSE deployments}
\label{tab:latency_avse}
\begin{tabular}{lll}
\toprule
\textbf{Network} & \textbf{Latency (ms)} & \textbf{Notes} \\
\midrule
Ethernet (Direct) 
& $\sim$5--15 
& Stable, lowest latency \\

5G private network (Cambridge Wireless) 
& $\sim$20--35 
& Satisfies real-time requirements \\

5G public network (Vodafone + AWS Wavelength, London) 
& 54.4--73 
& Acceptable for real-time A/V with reduced margin \\

Wi-Fi 4 (802.11n) 
& $\sim$50--100 
& Too high for smooth streaming; Wi-Fi 5/6 recommended \\

4G (Cambridge Wireless) 
& $\sim$60--150 
& Exceeds latency threshold, inconsistent \\

AWS over Internet (WebRTC) 
& $\sim$80--200 
& Region and ISP dependent; cloud adds delay \\
\bottomrule
\end{tabular}
\end{table*}


The latency measurements highlight the strong dependence of real-time audio-visual speech enhancement (AVSE) performance on network architecture rather than access technology alone. Ethernet (direct) provides the lowest and most stable latency ($\approx$5–15 ms), representing the ideal baseline for real-time AVSE and confirming that processing, rather than networking, becomes the dominant factor in wired deployments.

The private 5G network achieves consistently low latency ($\approx$20–35 ms), comfortably satisfying real-time requirements and demonstrating the advantage of localized cores and controlled radio environments for edge-based AVSE services.

Vodafone’s public 5G network with AWS Wavelength in London exhibits moderate latency ($\approx$54–73 ms), which remains acceptable for real-time audio-visual applications, though with reduced margin. This increase reflects the added complexity of public network traversal and edge-cloud integration, yet still significantly outperforms traditional internet-based cloud deployment.

In contrast, Wi-Fi 4 and 4G show higher and more variable latency ($\approx$50–150 ms), often exceeding thresholds for smooth interactive streaming. These results indicate that legacy access technologies and shared radio conditions introduce instability that limits AVSE usability.

Finally, deploying AVSE over the public internet to AWS regions using WebRTC results in the highest and most unpredictable latency ($\approx$80–200 ms), driven by geographic distance, ISP routing, and core network hops. This clearly demonstrates the importance of edge computing for latency-sensitive audio-visual applications.

Overall, the results confirm that edge proximity is the dominant factor enabling real-time AVSE, and that public 5G with edge cloud provides a practical middle ground between private networks and best-effort internet cloud deployments.

\subsection{Algorithm Processing Latency}

\begin{figure}
    \centering
    \includegraphics[width=\columnwidth]{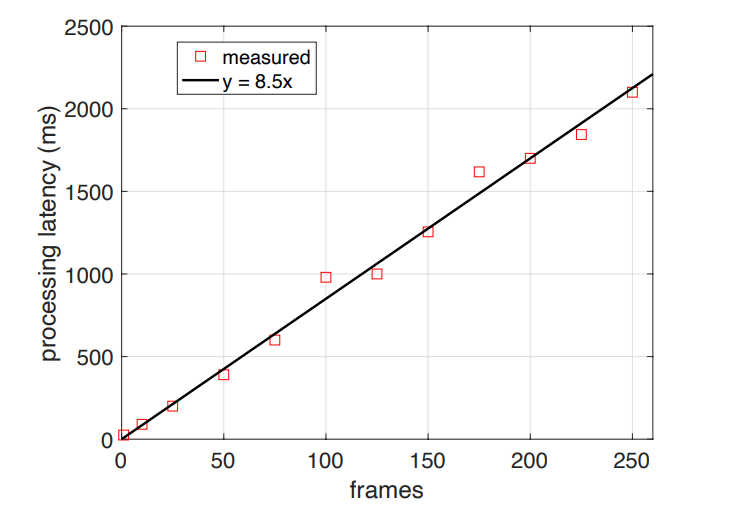}
    \caption{Algorithm processing latency versus input chunk size.}
    \label{figure6}
\end{figure}

The end-to-end user-perceived delay ($t_{\text{delay}}$) can be decomposed into communication latency ($t_{\text{comm}}$) and algorithm processing latency ($t_{\text{proc}}$), i.e.,
\begin{equation}
t_{\text{delay}} = t_{\text{comm}} + t_{\text{proc}}.
\end{equation}
In practice, $t_{\text{comm}}$ mainly affects the synchronization and coherence of the feedback audio, whereas $t_{\text{proc}}$ directly determines the system responsiveness. Real-world measurements indicate that modern wireless standards such as 5G NR and Wi-Fi 6 can satisfy the required throughput and transmission delay constraints. Consequently, minimizing $t_{\text{proc}}$ becomes the dominant factor in reducing overall interaction latency.

A common approach to reduce $t_{\text{proc}}$ is to shorten the input chunk duration and reduce model complexity. However, in audio--visual speech enhancement (AVSE), the enhancement model must capture temporal lip-motion dynamics in order to reliably extract the target speaker from competing noise sources. As a result, overly short input chunks may degrade enhancement performance due to insufficient temporal context.

This latency--performance trade-off has been widely recognized in the AVSE literature. Gogate \emph{et al.} introduced \textit{CochleaNet}, a causal, language-, noise-, and speaker-independent AVSE framework designed for real-time processing in noisy environments~\cite{gogate2020cochleanet}. More recently, Gogate \emph{et al.} proposed a robust real-time AVSE framework that integrates a CycleGAN-based visual enhancement stage with a causal DNN-based fusion model to estimate an ideal binary mask (IBM) for noise suppression~\cite{gogate2024robust}. Their system reports an end-to-end processing latency of approximately $15$~ms while achieving significant gains in objective and subjective intelligibility metrics, demonstrating the feasibility of low-latency AVSE under real-time constraints~\cite{gogate2024robust}. Additional efforts by the same group further explore lightweight architectures for benchmark AVSE challenges, emphasizing computational efficiency and practical deployment~\cite{gogate2024lightweight}. Furthermore, latency-efficient AVSE models based on cross-modal feature synchronization have reported real-time performance with processing delays on the order of $36$~ms, reinforcing the importance of low-latency multimodal fusion for assistive hearing applications~\cite{saleem2025audio}.

In our system, the input stream is segmented into chunks of $N$ frames, where each frame corresponds to $40$~ms. Fig.~\ref{figure6} illustrates the measured algorithm processing latency as a function of chunk size. The results indicate that processing latency increases approximately linearly with the input duration. In principle, processing each frame independently could reduce $t_{\text{proc}}$ below $40$~ms, enabling a total interaction latency within $100$~ms. However, such short-context processing significantly reduces enhancement quality due to the lack of sufficient temporal information for robust audio--visual fusion.

\begin{figure*}
    \centering
    \includegraphics[width=\textwidth]{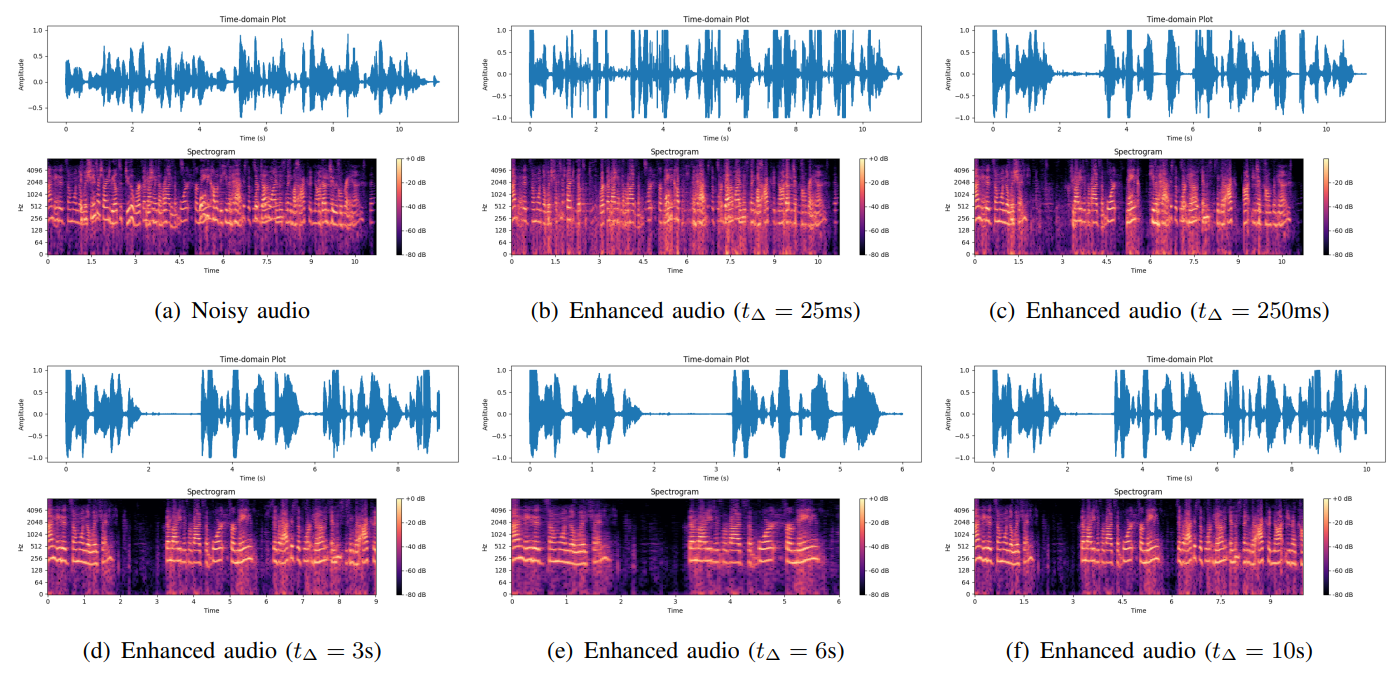}
    \caption{Time plot and spectrogram of the input noisy audio and output enhanced audio processed with different input chunk lengths.}
    \label{figure7}
\end{figure*}

To further evaluate the impact of chunk size on enhancement quality, Fig.~\ref{figure7} shows the time-domain waveform and spectrogram of the noisy input and enhanced output audio under varying chunk durations. Using $10$-s audio--visual segments produces the strongest reconstruction of the target speech and the most effective suppression of competing background noise. In contrast, frame-level enhancement yields output dominated by residual noise, indicating that the model fails to extract reliable speaker-discriminative features at this temporal resolution. Increasing the chunk duration generally improves enhancement quality, although the performance is not uniformly consistent across all temporal segments, suggesting that longer temporal context does not necessarily guarantee optimal reconstruction in all intervals.

\subsection{Algorithm Development}

To mitigate the processing latency while maintaining enhancement performance, we investigated whether the proposed model could learn discriminative representations from shorter temporal contexts. Specifically, the training scheme was adapted to support condensed input durations of $200$~ms. Based on the AVSE architecture shown in \autoref{fig:avse_arch}, a series of lightweight variants were developed by reducing network depth and neuron count with the objective of shortening inference time. The total number of parameters and measured execution time for each reduced model are summarized in \autoref{tab:table1}.

\begin{table}[]
    \centering
    \caption{Model Complexity and Execution Time}
    \begin{tabular}{|c|c|c|}
    \hline
        \textbf{Model} & \textbf{Total Parameters} &  \textbf{Execution Time (ms)} \\
        \hline 
        Model 1 &  1,540,396 (5.88 MB) &  1200 \\
        Model 2 &  603,564 (2.30 MB) &  550 \\
        Model 3 &  202,564 (791.27 KB) &  350 \\
        \hline
    \end{tabular}
    \label{tab:table1}
\end{table}

\begin{figure*}
    \centering
    \includegraphics[width=\textwidth]{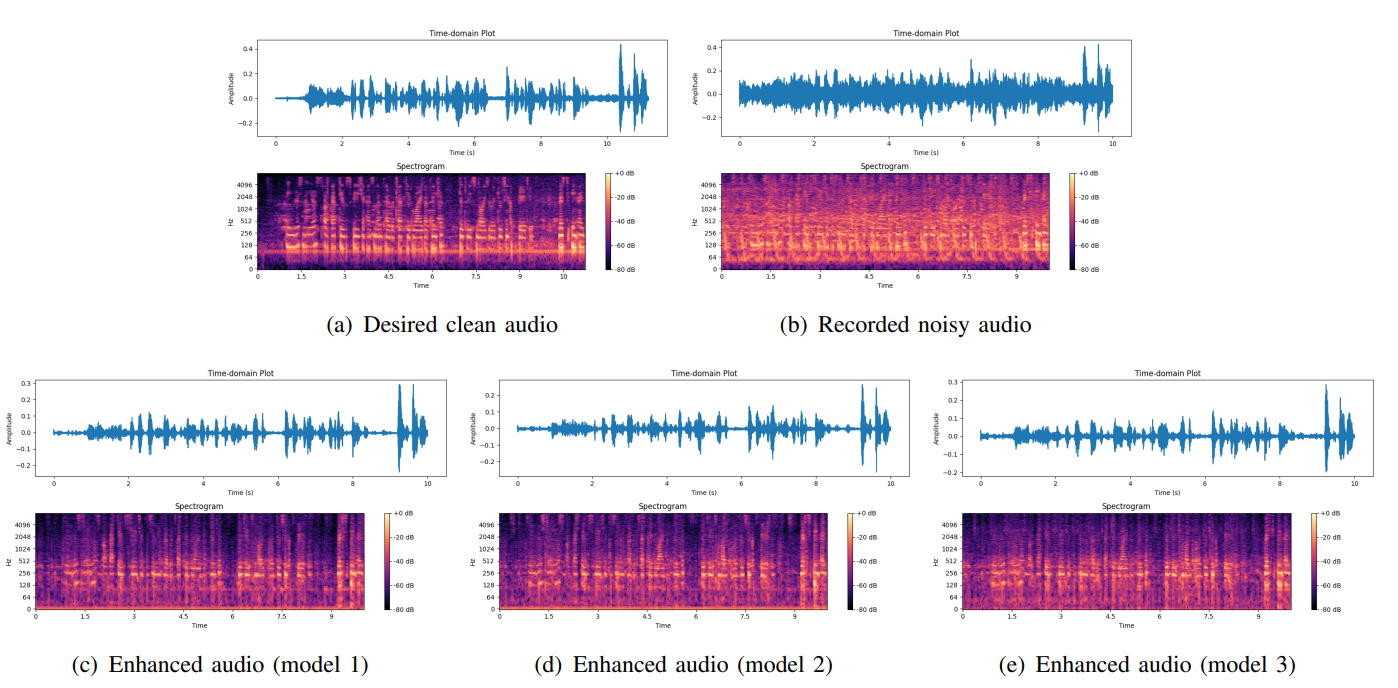}
    \caption{Time plot and spectrogram of the input noisy audio and output enhanced audio processed with different network architectures.}
    \label{figure8}
\end{figure*}

Fig.~\ref{figure8} compares the time-domain waveform and spectrogram of the clean target speech, noisy mixture, and enhanced output produced by the reduced models. The results show that shortening the input duration and reducing model capacity decreases reconstruction fidelity, particularly for low-energy speech components, due to reduced availability of visual temporal information. Nevertheless, the reduced models maintain satisfactory noise suppression performance, indicating that lightweight architectures can provide an effective trade-off between enhancement quality and low-latency execution.

Overall, the results demonstrate a clear trade-off between processing latency and enhancement performance. Longer input chunks provide richer temporal context for lip-motion tracking and speaker discrimination, but significantly increase processing delay. Conversely, reducing chunk size and model complexity improves responsiveness but may degrade speech reconstruction quality in challenging low-SNR conditions. Therefore, selecting an optimal configuration requires balancing the latency constraints of real-time feedback against the temporal context required for robust AVSE.

\section{CONCLUSION}
\label{Conclusion}

This work presented a comprehensive evaluation of a real-time, cloud–edge–assisted audio-visual speech enhancement (AVSE) system operating over a public 5G edge cloud. The proposed system integrates CNN-based acoustic processing and OpenCV-driven facial feature extraction with an LSTM fusion network to preserve temporal coherence, and was deployed and evaluated using a Vodafone-compatible AWS Wavelength edge in London. The results demonstrate that compute placement at the network edge and network transport latency are primary determinants of end-to-end performance for interactive multimedia and perceptual enhancement applications.

Experimental analysis showed that real-time coherence is achieved only when communication delay remains below the chunk duration ($t_{\text{comm}} \leq t_{\text{chunk}} = 40$ ms). In practice, this requirement was consistently met by 5G and wired Ethernet connections for uncompressed 0.3 MB audio-video chunks, while other access networks failed to maintain stable operation. Applying aggressive compression (up to a factor of 80) reduced payloads to approximately 70 KB with negligible perceptual degradation, significantly improving robustness under constrained uplink conditions and enabling smoother real-time streaming. These findings confirm that uplink capacity is often the dominant bottleneck for interactive edge-based multimedia services.

We further observed that system responsiveness and enhancement quality are jointly governed by network quality, chunk size, and model complexity. Reducing temporal context and model size lowered processing latency from approximately 1.2 s to 0.35 s, but at the cost of degraded speech reconstruction, particularly in low-SNR conditions. This reveals a fundamental trade-off between computational efficiency and enhancement fidelity, underscoring the need for adaptive operating profiles that can dynamically balance latency, throughput, and quality based on current network and compute conditions. Stress testing identified safe operating regions as well as the importance of fallback modes, such as audio-only operation, to preserve service continuity during adverse network conditions.

Overall, the results indicate that public 5G edge environments can sustain real-time, interactive multimedia workloads, provided that network and compute resources are carefully orchestrated. However, performance margins remain tighter than in dedicated or private infrastructures, making dynamic adaptation and edge proximity essential architectural considerations. The insights derived from this study are directly applicable to delay-sensitive services such as augmented reality, interactive conferencing, and multimodal perception enhancement.

Future work will extend this study by incorporating mobility and multi-user scenarios, exploring GPU-accelerated edge instances for more complex models, and evaluating performance across multiple geographic regions and operator ecosystems. We also plan to investigate adaptive chunking, lightweight streaming architectures, and edge-assisted learning to further reduce latency while preserving enhancement fidelity in real-world wireless environments.

\section*{Acknowledgements}
This work was supported by the UK Engineering and Physical Sciences Research Council (EPSRC) Grant No. EP/T021063/1. 
\printbibliography

@inproceedings{gupta20235g,
  title={5G-IoT Cloud based Demonstration of Real-Time Audio-Visual Speech Enhancement for Multimodal Hearing-aids},
  author={Gupta, Ankit and Bishnu, Abhijeet and Gogate, Mandar and Dashtipour, Kia and Arslan, Tughrul and Adeel, Ahsan and Hussain, Amir and Ratnarajah, Tharmalingam and Sellathurai, Mathini},
  booktitle={24th International Speech Communication Association Conference 2023},
  pages={686--687},
  year={2023}
}

@article{gogate2020cochleanet,
  title={CochleaNet: A robust language-independent audio-visual model for real-time speech enhancement},
  author={Gogate, Mandar and Dashtipour, Kia and Adeel, Ahsan and Hussain, Amir},
  journal={Information Fusion},
  volume={63},
  pages={273--285},
  year={2020},
  publisher={Elsevier}
}

@article{gogate2024robust,
  title={Robust real-time audio-visual speech enhancement based on dnn and gan},
  author={Gogate, Mandar and Dashtipour, Kia and Hussain, Amir},
  journal={IEEE Transactions on Artificial Intelligence},
  year={2024},
  publisher={IEEE}
}

@article{gogate2024lightweight,
  title={A lightweight real-time audio-visual speech enhancement framework},
  author={Gogate, Mandar and Dashtipour, Kia and Hussain, Amir},
  journal={Proceedings of AVSEC},
  pages={19--23},
  year={2024}
}

@article{saleem2025audio,
  title={Audio-visual feature synchronization for robust speech enhancement in hearing aids},
  author={Saleem, Nasir and Gogate, Mandar and Dashtipour, Kia and Hussain, Adeel and Anwar, Usman and Adetomi, Adewale and Arslan, Tughrul and Hussain, Amir},
  journal={arXiv preprint arXiv:2508.19483},
  year={2025}
}

@misc{etsi_mec,
  title        = {Multi-access Edge Computing (MEC)},
  author       = {{European Telecommunications Standards Institute (ETSI)}},
  howpublished = {\url{https://www.etsi.org/technologies/multi-access-edge-computing}},
  note         = {Accessed: 2026-01},
}

@misc{aws_wavelength,
  title        = {5G Edge Computing Infrastructure – AWS Wavelength},
  author       = {{Amazon Web Services, Inc.}},
  howpublished = {\url{https://aws.amazon.com/wavelength/}},
  note         = {Accessed: Jan. 2026},
}

@article{hou2018audio,
  title={Audio-visual speech enhancement using multimodal deep convolutional neural networks},
  author={Hou, Jen-Cheng and Wang, Syu-Siang and Lai, Ying-Hui and Tsao, Yu and Chang, Hsiu-Wen and Wang, Hsin-Min},
  journal={IEEE Transactions on Emerging Topics in Computational Intelligence},
  volume={2},
  number={2},
  pages={117--128},
  year={2018},
  publisher={IEEE}
}

@article{ephrat2018looking,
  title={Looking to listen at the cocktail party: A speaker-independent audio-visual model for speech separation},
  author={Ephrat, Ariel and Mosseri, Inbar and Lang, Oran and Dekel, Tali and Wilson, Kevin and Hassidim, Avinatan and Freeman, William T and Rubinstein, Michael},
  journal={arXiv preprint arXiv:1804.03619},
  year={2018}
}

@inproceedings{zhu2023real,
  title={Real-time audio-visual end-to-end speech enhancement},
  author={Zhu, Zirun and Yang, Hemin and Tang, Min and Yang, Ziyi and Eskimez, Sefik Emre and Wang, Huaming},
  booktitle={ICASSP 2023-2023 IEEE International Conference on Acoustics, Speech and Signal Processing (ICASSP)},
  pages={1--5},
  year={2023},
  organization={IEEE}
}

@article{adeel2018real,
  title={Real-time lightweight chaotic encryption for 5g iot enabled lip-reading driven secure hearing-aid},
  author={Adeel, Ahsan and Ahmad, Jawad and Hussain, Amir},
  journal={arXiv preprint arXiv:1809.04966},
  year={2018}
}

@article{mao2017survey,
  title={A survey on mobile edge computing: The communication perspective},
  author={Mao, Yuyi and You, Changsheng and Zhang, Jun and Huang, Kaibin and Letaief, Khaled B},
  journal={IEEE communications surveys \& tutorials},
  volume={19},
  number={4},
  pages={2322--2358},
  year={2017},
  publisher={IEEE}
}

@article{yuan2022network,
  title={Network-aware 5G edge computing for object detection: Augmenting wearables to “see” more, farther and faster},
  author={Yuan, Zhongzheng and Azzino, Tommy and Hao, Yu and Lyu, Yixuan and Pei, Haoyang and Boldini, Alain and Mezzavilla, Marco and Beheshti, Mahya and Porfiri, Maurizio and Hudson, Todd E and others},
  journal={IEEE Access},
  volume={10},
  pages={29612--29632},
  year={2022},
  publisher={IEEE}
}
\end{document}